%% file: ms.tex
\documentclass[journal]{IEEEtran}
\usepackage{multirow}
\usepackage{times} 
\usepackage{mathtools}
\usepackage{scalerel,amssymb}     
\usepackage{xcolor}
\usepackage{cite}
\usepackage{bm}
\usepackage{graphics}
\graphicspath{ {./Images/} }
\usepackage[font={small}]{caption}

\usepackage{color}
\usepackage{amsmath}
\usepackage{amsfonts}
\usepackage{titlesec}
\usepackage{indentfirst}
\usepackage{textcomp}
\usepackage{footnote}
\usepackage[flushleft]{threeparttable} %
\usepackage{adjustbox}
\usepackage{booktabs}

\usepackage{textcomp}
\usepackage{array, makecell} %
\usepackage{lettrine}
\usepackage[caption=false]{subfig}
\usepackage{siunitx}
\usepackage{color, colortbl}
\definecolor{Gray}{gray}{0.9}
\definecolor{BlueGreen}{rgb}{0.051,0.596,0.72}

\title
{
\LARGE \bf
Quantitative Physical Ergonomics Assessment of \\Teleoperation Interfaces
}

\author
{
Soheil Gholami$^{1,2}$, Marta Lorenzini$^{1}$, Elena De Momi$^{2}$, and Arash Ajoudani$^{1}$
\thanks{This work was supported by the European Research Council (ERC) starting grant Ergo-Lean, Grant Agreement No. 850932}
\thanks{$^{1}$ Human-Robot Interfaces and physical Interaction (HRI$^2$) Lab, Istituto  Italiano di Tecnologia, Genoa, Italy, 
Email: {\tt\small soheil.gholami@iit.it}, {\tt\small marta.lorenzini@iit.it}}
\thanks{$^{2}$ NearLab, Dept. of Electronics, Information and Bioengineering, Politecnico di Milano, Milan, Italy}
}%

\begin{document}

\maketitle
\thispagestyle{empty}
\pagestyle{empty}

\input{Sections/abstract}
\input{Sections/introduction}

\input{Sections/methodology}
\input{Sections/exp_setup}
\input{Sections/exp_validation}

\input{Sections/conclusions}

\bibliographystyle{IEEEtran}
\bibliography{biblio.bib}

\end{document}

%% file: Sections/abstract.tex
\begin{abstract}
Human factors and ergonomics are the essential constituents of teleoperation interfaces, which can significantly affect the human operator's performance. Thus, a quantitative evaluation of these elements and the ability to establish reliable comparison bases for different teleoperation interfaces are the keys to select the most suitable one for a particular application. However, most of the works on teleoperation have so far focused on the stability analysis and the transparency improvement of these systems, and do not cover the important usability aspects. In this work, we propose a foundation to build a general framework for the analysis of human factors and ergonomics in employing diverse teleoperation interfaces. The proposed framework will go beyond the traditional subjective analyses of usability by complementing it with online measurements of the human body configurations. As a result, multiple quantitative metrics such as joints' usage, range of motion comfort, center of mass divergence, and posture comfort are introduced. To demonstrate the potential of the proposed framework, two different teleoperation interfaces are considered, and real-world experiments with eleven participants performing a simulated industrial remote pick-and-place task are conducted. The quantitative results of this analysis are provided, and compared with subjective questionnaires, illustrating the effectiveness of the proposed framework.
\end{abstract}


%% file: Sections/introduction.tex
\section{Introduction}
\label{section:introduction}
\IEEEPARstart{T}{elerobotics}
 is of significant importance in applications
where the tasks are dangerous or even inaccessible to humans. So far, the related theoretical and experimental research studies have mainly focused on the overall system's safety (e.g., stability \cite{santiago2017stable}), tracking problem (e.g., transparency \cite{milstein2018human}), interaction uncertainties (e.g., tele-impedance regulation \cite{ajoudani2012tele}), and shared autonomy \cite{lin2020shared}. 

As a result, numerous teleoperation user-interfaces (UIs) have been presented \cite{adamides2014usability,shahbazi2018systematic}. 
This diversity, however, makes it difficult for the system designers to adopt a suitable UI for a specific application, which optimises the requirements of the target tasks. Hence, supportive metrics to evaluate and compare the advantages and disadvantages of these UIs become necessary. Nevertheless, to the best of our knowledge, this problem has received very little attention in the literature. To respond to this need, the usability aspects of the teleoperation systems must be analysed through the definition of a set of ergonomic and performance indexes.

In telerobotics, ergonomics is investigated within the physical and cognitive aspects, which may influence each other as well \cite{mehta2016integrating}. Roughly speaking, the user's motion generation interface contributes more to the physical-related factors while the perception feedback plays a significant role in the cognitive ones. Ergonomics, however, is not the only design factor for teleoperation systems. A trade-off must be made between ergonomic aspects and performance measures such as execution time, learning-curve status, success rate, and the similarity between the human and robot motions.

 \begin{figure}[!t]
     \centering
         \centering
         \includegraphics[trim=0.1cm 0.5cm 0.1cm 0.1cm,clip, width=0.95\columnwidth]{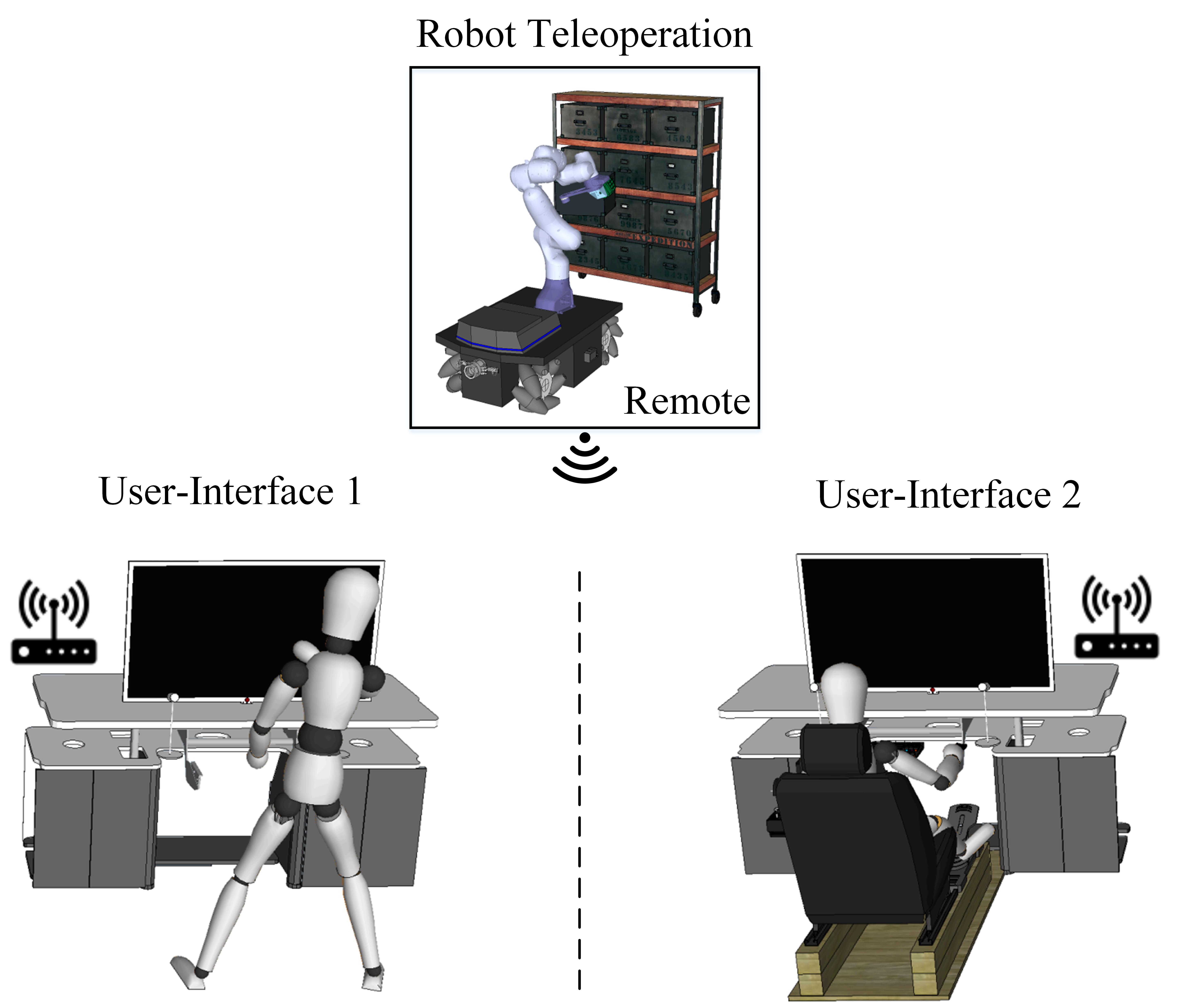}
         \caption{The conceptual illustration of a user teleoperating a remote collaborative robot with two different user-interfaces: (i) standing interface with a whole-body motion capture system and (ii) seated interface with a 3D mouse.}
         \label{fig:into}
\end{figure}

In this work, we propose a set of quantitative metrics to assess the teleoperation UIs in terms of human factors and ergonomics. In developing such metrics, we inspire from the ``postures and movement" section of the Ergonomic Assessment Work-Sheet (EAWS) \cite{schaub2013european,bortolini2020motion}. To be consistent with EAWS, we adopt a systematic process to appraise both the upper and lower parts of the human musculoskeletal system. 
The scoring system is based on the pre-defined standard body postures, accepted joints' Range of Motion (RoM), body features and dimensions, and the workstation layout. 
In addition, the qualitative subjective analyses on time and cost efficiencies, and usability are carried out. These enable the designers to choose an appropriate teleoperation UI for a specific application, regarding the body kinematic information and the task's performance metrics. Another use-case of the developed framework is the optimization of the teleoperation workstation layout by reorganizing the geometrical properties of a UI such as the monitor-display height. 

The proposed framework is evaluated through quantitative and qualitative experimental analyses on eleven healthy subjects. Two popular teleoperation interfaces are considered here: (i) standing and (ii) seated. The latter uses a 3D mouse device while, in the former, a Motion Capture system (MoCap) is employed. 
Another role of this MoCap system is to capture the human kinematic data needed by the introduced ergonomics indexes during the investigation of the afore-mentioned teleoperation UIs. For the follower system, the MObile Collaborative Robotic Assistant (MOCA) \cite{MocaTeleop} is utilised as the follower robot, controlled in two sepaprate control modes, i.e., locomotion and manipulation modes. Conceptual illustration of a user teleoperating the MOCA with the afore-mentioned UIs is shown in Fig. \ref{fig:into}.

The choice for the two studied teleoperation interfaces was made based on their inherently different characteristics, in terms of dexterity and comfort. Hence, if the proposed indexes can be quantified in these two ``extreme" settings, several existing UIs that implement a trade off between teleoperation dexterity and user comfort (e.g., UIs based on haptic devices)  are also likely to be quantified using the proposed approach.

\section{Related work}
\label{section:related_work}
Most of the research studies about the ergonomics assessment of teleoperation systems have been devoted to the effects of time delays on the human cognitive (mental) workload. For instance, the relation between the time delays and human factors, situational awareness, and cognitive load was analysed for both local and remote assembly tasks in \cite{kumar2020methodology}. A similar approach was presented in \cite{rogers2017investigation} to study the outcomes of latency on the performance, subjective workload, trust, and usability measures for manual multi-robot teleoperated tasks. 
Moreover, a qualitative usability evaluation was carried out in \cite{adamides2017hri}, for a scenario in which an agricultural robotic sprayer was used in a teleoperation setup.
The authors considered three factors in their experiments: (i) input devices (PC keyboard and SONY PlayStation{\textregistered} game-pad), (ii) output devices (PC screen and head-mounted display),
and (iii) number of camera views (single and multiple viewpoints). They studied the effects of these factors on the operator's workload and awareness status.
However, the evaluation was performed by means of the so-called subjective questionnaires only.   

On the other hand, less attention has been paid to the physical ergonomic aspects.
For instance, in \cite{lin2019physical}, the authors carried out a user study to investigate the physical fatigue of the operators, while they teleoperated a mobile humanoid robot. Nevertheless, just one motion generation interface (whole-body optical MoCap system) was considered. In addition, the proposed assessment technique was based on surface ElectroMyoGraphy (sEMG) sensors. As it is well known, the sEMG signals are highly contaminated with noises and artifacts and thus their interpretation is questionable, especially in the case of dynamic and rapid motions. Besides, an inverse kinematic model of the human arm and the Rapid Upper Limb Assessment (RULA) metric were utilised in a haptic-enabled shared control teleoperation approach to estimate the current user’s comfort online in \cite{rahal2020caring}. They reported a $30\%$ perceived reduction of the workload with respect to the pure teleoperation case. Nonetheless, the whole-body ergonomics evaluation was not considered and the authors just used one particular teleoperation UI in their study.

For the \textit{local direct} Human-Robot Interaction and Collaboration (HRI/C), however, the problem of the physical ergonomics assessment has been widely studied in the literature. As an example, the authors in \cite{kim2019adaptable} developed a framework to realise a reconfigurable HRI/C system, which aimed at improving the workers' ergonomics and productivity in manufacturing environments. This was accomplished by the online perception of the human states (body kinematics and dynamics), tools, and the environment. Consequently, the robot adapted itself based on the optimality of the introduced metrics for human factors. A set of required mental and physical factors for human comfort during the cooperation with robotic systems in a shared area was also investigated in \cite{changizi2018comfort}. 

Moreover, some efforts have been made in an attempt to define performance metrics for the HRI/C setups. Still, the comparison aspects for evaluation purposes of the teleoperation UIs were missed. For example, in \cite{son2010enhancement}, the authors proposed the surgeons' kinaesthetic perception and the position-tracking ability of leader-follower system as performance metrics. These were utilised to enhance the kinaesthetic perception condition of the task while maintaining the system's stability and tracking requirements. Indeed, this trade-off was solved by means of a multi-constrained optimisation approach. Apart from kinematic metrics, a set of quantitative physiological and task performance metrics were introduced in \cite{wang2018toward}. They studied the effects of teleoperation motion mappings (joint-space control etc.) on these metrics in a robotic needle steering scenario. Nonetheless, these were just evaluated with one teleoperation UI, i.e., the haptic device. Besides, some quantitative performance and awareness metrics were proposed in \cite{hong2017multimodal} for the multiple mobile-robot scenarios. This work aimed to enhance the situational awareness of human operators in the remote outdoor navigational tasks, and the manipulation mode was not studied. Moreover, in \cite{buzzi2018biomimetic}, the authors utilised a novel biomimetic impedance modulation controller to improve the task-execution accuracy in a virtual targeting task. Consequently, they showed an increase in the users’ performance in terms of positional error and overshoots from the targets. This was done by introducing multiple quantitative performance indexes based on the sEMG signals without evaluating the effects of different interface setups.

%% file: Sections/methodology.tex
\section{Framework Definition}
\label{section:framework_def}
\begin{figure}[!t]
     \centering
     \includegraphics[trim=0cm 0cm 0cm 0cm, width=0.4\textwidth]{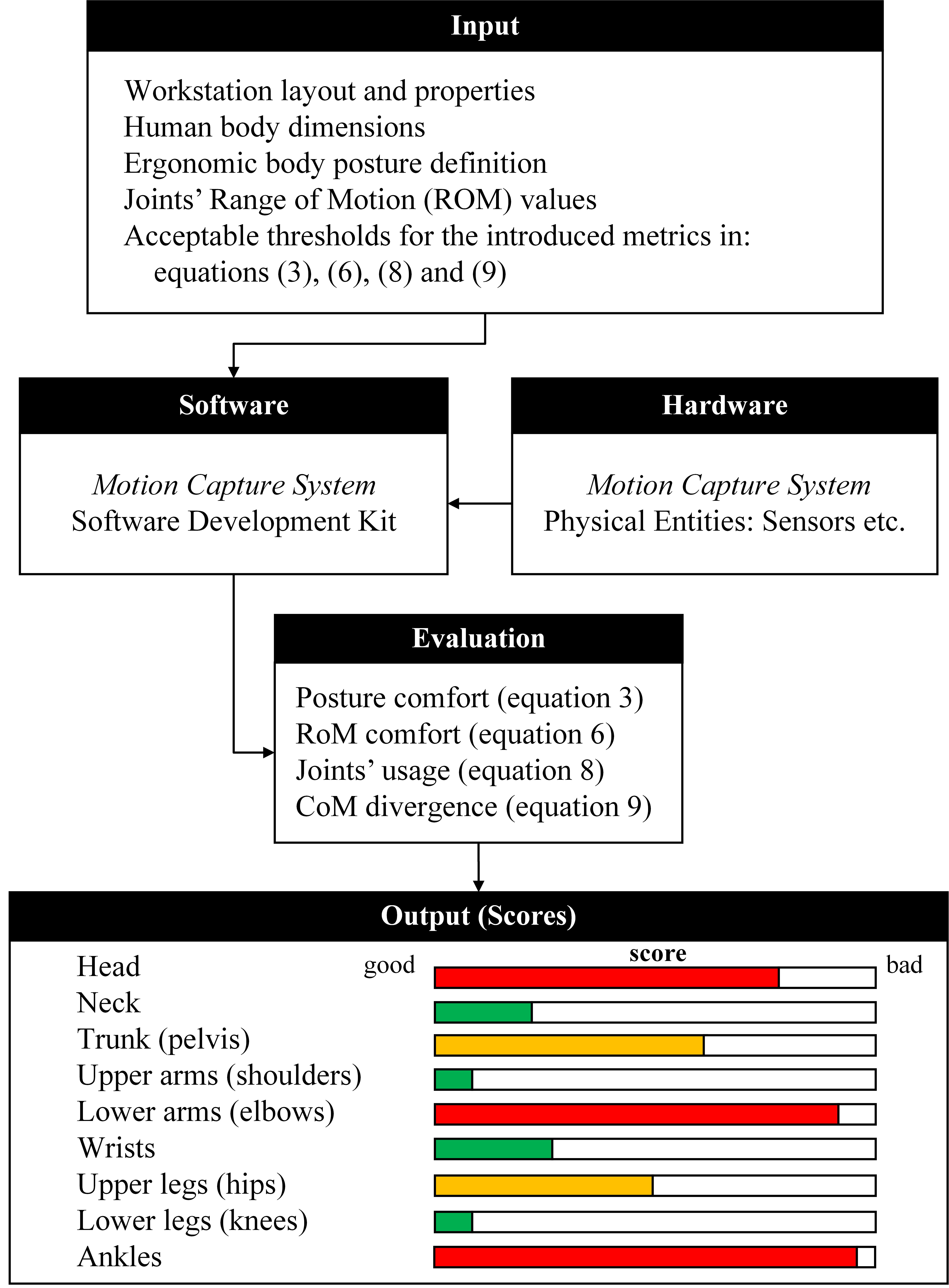}
     \caption{The block diagram of the proposed physical ergonomics assessment framework for teleoperation UIs. Regarding the percentage values in the ``output (scores)" block, these are symbolic representations based on the defined acceptable thresholds and the values of equations \eqref{metric_joints_usage}, \eqref{metric_rom}, \eqref{metric_cog}, and \eqref{optimal_pose} at each time instant.
    }
    \label{fig:block_diagram}
\end{figure}

The general structure of the proposed framework is illustrated in Fig. {\ref{fig:block_diagram}}. In the ``input" block, the task's parameters such as workstation layout and properties, human body dimensions, ergonomic body posture definition, joints' RoM values, and the metrics' acceptable thresholds are defined.
The ``hardware" block includes the MoCap's physical entities such as sensors and signal transmitter/receiver modules. The only requirement for this part is the capability of online measurement of the body joints' states. The corresponding measurements are, then, analysed in the ``software" block. Indeed, the received raw data is processed, and the relative human body kinematic information is generated in the related Software Development Kit (SDK) of the employed MoCap system.
As a result, the processed data is read in the ``evaluation" block to calculate the proposed kinematic metrics through equations \eqref{metric_joints_usage}, \eqref{metric_rom}, \eqref{metric_cog}, and \eqref{optimal_pose}. After this step, the normalized values are utilised in the ``output (scores)" panel to score the essential body links and joints. 

Moreover, a set of offline measures is introduced in Section \ref{perf_metric}, as the complementary qualitative measures to the quantitative ones (Section \ref{section:metrics}). 
In particular, the National Aeronautics and Space Administration Task Load Index (NASA-TLX) tool \cite{hart1986nasa} is selected to estimate the perceived workload of each studied teleoperation UI. 
This can also be used as a verification tool for the developed quantitative metrics.
\subsection{Kinematics-related ergonomics metrics}
\label{section:metrics}
The musculoskeletal system of the human body can be modelled by $n$ articulations (joints) and $n$ body segments (links) \cite{myn2015xsens}. The conceptual illustration of this parent-child structure with the definition of the joints' frames $\Sigma_{i}$  ($i \in \{1,\,\dots,\,n\}$) and the world frame $\Sigma_W$ is shown in Fig.~\ref{fig:body}-a. 
The $i$-th joint's position (angle), velocity, and acceleration with respect to (w.r.t.) its parent link are denoted with $\bm{\theta}_i\in \mathbb{R}^{n_i}$, $\bm{\dot{\theta}}_i$, and $\bm{\ddot{\theta}}_i$, respectively. 
$n_i \leq 3$ is the number of $i$-th joint's Degrees-of-Freedom (DoFs). Each DoF imposes a set of constraints on its subsequent link's motion patterns, which leads to the definition of the following joint's RoM:
\begin{equation}
\small
    \label{ROM}
    {\bm{\theta}_i^j}_{min} < \bm{\theta}_i^j < {\bm{\theta}_i^j}_{max},\, i \in \{1,\,\dots,\,n\},\,j \in \{1,\,\dots,\,n_i\}.
\end{equation}
Besides, the pose (position $\bm{p}_i = {[x_i,\,y_i,\,z_i]}^T \in \mathbb{R}^3$ and orientation $\bm{\epsilon}_i = {[q_1,\,q_2,\,q_3,\,q_4]}^T \in \mathbb{R}^4$)
of each link w.r.t. the world frame is defined with $\bm{x}_i = {[\bm{p}^T_i,\, \bm{\epsilon}^T_i]}^T \in \mathbb{R}^{7}$. It should be noted that the orientation is represented by the quaternions.

In what follows, the proposed kinematics-related ergonomics metrics, which are defined for the simplified musculoskeletal system (Fig. \ref{fig:body}-b), are introduced.

\subsubsection{Posture comfort}
Assume that $\bm{x}_i = {[\bm{p}_i,\,\bm{\epsilon}_i]}^T$ is the current pose of $\Sigma_i$ w.r.t. $\Sigma_W$. Also, the comfortable (ergonomic) pose of each link is denoted by ${\bm{x}^{\star}}_i = {[\bm{p}_i^{\star},\,\bm{\epsilon}_i^{\star}]}^T$. So,
the link's position displacement $\rho_i$ and its quaternion displacement $\eta_i$ w.r.t. to the comfortable pose are defined as follows:
\begin{equation}
\label{psture_comfort_8}
    \rho_i(k) =
    {\left\| 
    \bm{p}^T_i(k) - {\bm{p}_i^{\star}}^T
    \right\|}_{\mathcal{L}_2}, \quad
    \eta_i(k) =
    d\left(
    \bm{\epsilon}_i(k),\, \bm{\epsilon}_i^{\star}
   \right),
\end{equation}
where $d(\hat{\bm{\epsilon}}_1,\,\hat{\bm{\epsilon}}_2 ) \triangleq 1 - {\langle \hat{\bm{\epsilon}}_1,\,\hat{\bm{\epsilon}}_2\rangle}^2 $, with ${\langle \hat{\bm{\epsilon}}_1,\,\hat{\bm{\epsilon}}_2\rangle}$ being the inner  product of two unit quaternions $\hat{\bm{\epsilon}}_1$ and $\hat{\bm{\epsilon}}_2$. This quantity becomes $0$ whenever the unit quaternions represent the same orientation.
Equation \eqref{psture_comfort_8} leads to  definition of the following metric $\zeta_i$ for checking the posture status at time instant $k$:
\begin{equation}
\label{optimal_pose}
   \zeta_i(k)  \triangleq \rho_i(k) +  {w_{\eta}}\,\eta_i(k),
\end{equation}
where ${w_{\eta}} \in \mathbb{R}$  is the orientation scaling factor and is set to $\pi$ (it maps the values from $[0,\,1]$ to $[0,\,\pi]$).

Regarding the ergonomic body posture, it is the one with the least generated muscular efforts. Up to now, several ideal postures are defined for different workstation layout. For instance, one can refer to the neutral body posture, the posture the human body naturally assumes in micro-gravity \cite{han2019neutral}, or the standard ergonomic postures presented as guidelines for different workstations \cite{pheasant2005bodyspace,mckenzie7steps,boulila2018ergonomics}. In this work, the latter is used as our standard reference posture. Hence, adopting any other posture requires a non-optimal amount of muscular efforts that can cause fatigue or even musculoskeletal disorders during a prolonged task. 
\subsubsection{RoM comfort}
Based on the joints' RoM values \eqref{ROM}, the following comfort quantity $\bm{\xi}^j_i$ can be defined for the $j$-th DoF of the $i$-th joint at time instant $k$, similar to the approach suggested in \cite{moschonas2011novel}:
\begin{equation}
    \bm{\xi}^j_i(k) \triangleq \min
    \left\{
    |{\bm{\theta}^j_i}(k) - {\bm{\theta}^j_i}_{min}|,\,
    |{\bm{\theta}^j_i}(k) - {\bm{\theta}^j_i}_{max}|
    \right\},
\end{equation}
which can be normalized to the interval of $[0,\,1]$ by:
\begin{equation}
    {\bm{\xi}^j_i}^{\dagger}(k) = \frac{2\,\bm{\xi}^j_i(k)}
    {|{\bm{\theta}^j_i}_{max} - {\bm{\theta}^j_i}_{min}|}.
\end{equation}
The more closer ${\bm{\xi}^j_i}^{\dagger}$ is to $1.0$, the more comfortable posture the user experiences. After summing up the effects of $n_i$ DoFs of the $i$-th joint, we have:
\begin{equation}
\label{metric_rom}
    {{\bar{\bm{\xi}}^{\dagger}_i}}(k) = 
    \sum_{j=1}^{n_i} 
    {\bm{\xi}^j_i}^{\dagger}(k).
\end{equation}
\subsubsection{Joints' usage}
For this metric we assume that the more variations in the joints' position values in time, the more complex, thus demanding, the task can be classified.
To track the changes of the position of $j$-th DoF of the $i$-th joint at time instant $k$, we define two metrics: 
(i) $\bar{\bm{\psi}}^j_i$ - absolute divergence from the average value and 
(ii) ${\bm{\psi}}^j_i$ - absolute divergence from the previous value. These are: 
\begin{equation}
    \bar{\bm{\psi}}^j_i(k) = {\left|{\bm{\theta}^j_i}(k) - {\bar{\bm{\theta}}^j_i}\right|},  \,\,
    \bm{\psi}^j_i(k) = {\left|{\bm{\theta}^j_i}(k) - {{\bm{\theta}}^j_i(k-1)}\right|},
\end{equation}
where, ${\bar{\bm{\theta}}^j_i}$
is the average value of ${\bm{\theta}^j_i}$ in the duration of interest. So, 
for each joint we have:
\begin{equation}
\label{metric_joints_usage}
\bar{\bm{\psi}}_i(k) = 
\sum_{j=1}^{n_i} 
\bar{\bm{\psi}}^j_i(k),\quad
{\bm{\psi}}_i(k) = 
\sum_{j=1}^{n_i} 
{\bm{\psi}}^j_i(k).
\end{equation}
\subsubsection{CoM divergence}
\begin{figure}[!t]
         \centering
         \includegraphics[trim=0.25cm 0.5cm 0.25cm 0.25cm, width=0.975\linewidth]{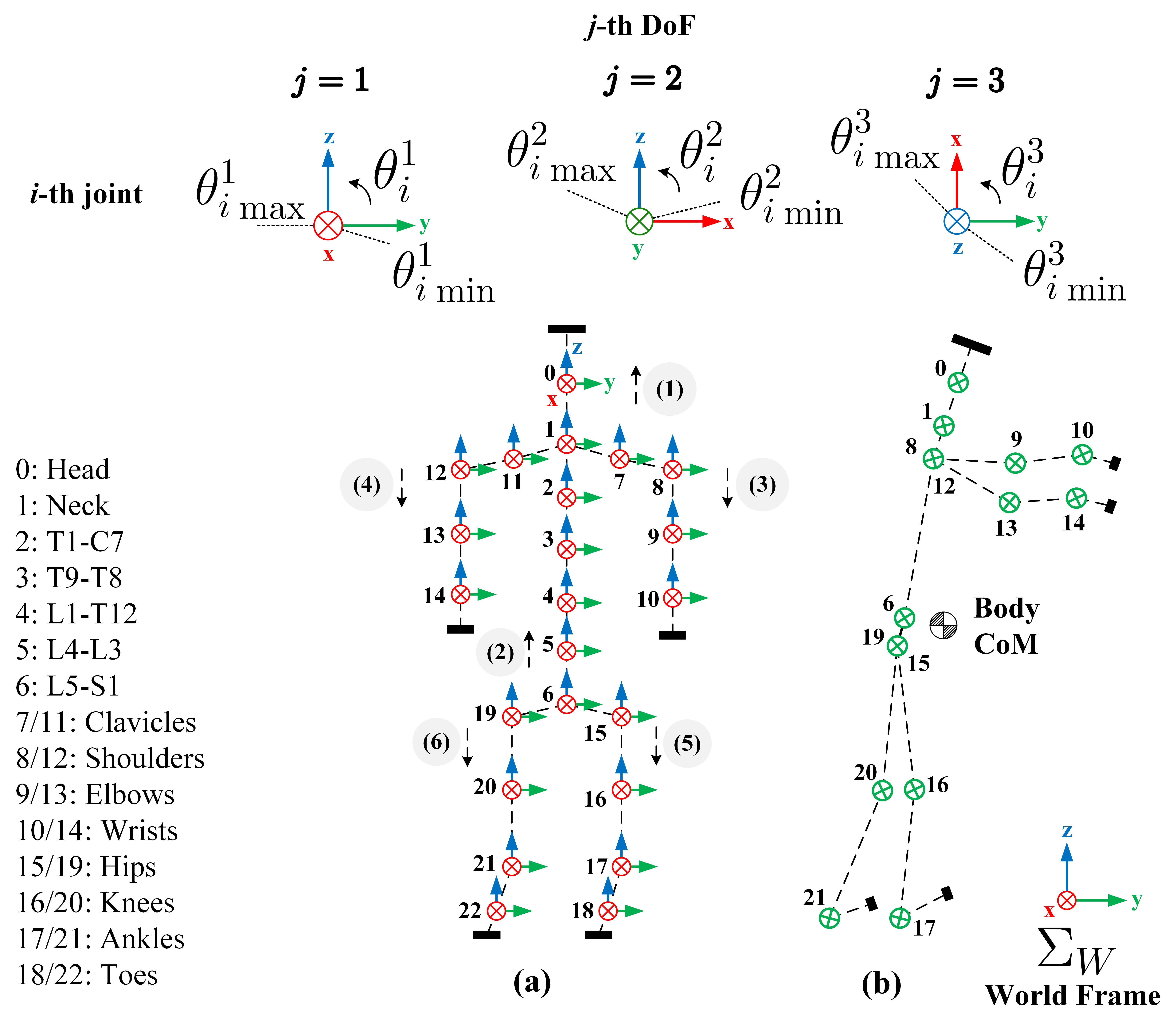}
         \caption{
         {(a) Musculoskeletal system of the human body in Neutral pose (N-pose). (b) Simplified model considering the essential joints (side-view). CoM stands for Center of Mass. Joints and links are represented by $\otimes$ symbol and dashed lines, respectively. The numbered dotted arrows show the chains' directions. To avoid clutter, the joint's number is also used to show the  joints' frames $\Sigma_i$. 
         }
         }
         \label{fig:body}
\end{figure}

Changes in the position of the body CoM (Fig. \ref{fig:body}-b) 
$\bm{p}^T_{CoM} = [\bm{p}^x_{CoM},\,\bm{p}^y_{CoM},\,\bm{p}^z_{CoM}]$ 
reveal the body translational motions indicating the engagement level of the user's whole-body in the task. Proposed method in \cite {floor2012use} can be used for CoM estimation. Regarding this quantity, we use the 
$\mathcal{L}_2$-norm 
measure to calculate the divergence of 
$\bm{p}^T_{CoM}$ 
from its average point 
$\bar{\bm{p}}^T_{CoM}$ 
and its previous value 
${\bm{p}}^T_{CoM}(k-1)$. We have:
\begin{equation}
\label{metric_cog}
\begin{split}
    \delta {p}^{\star}_{CoM}(k) &= 
    {\left\|\bm{p}^T_{CoM}(k) - \bar{\bm{p}}^T_{CoM}\right\|}_{\mathcal{L}_2}, \\
    \delta {p}_{CoM}(k) &= 
    {\left\|\bm{p}^T_{CoM}(k) - \bm{p}^T_{CoM}(k-1)\right\|}_{\mathcal{L}_2}.
\end{split}
\end{equation}
\subsection{Qualitative measures and performance metrics}
\label{perf_metric}
Assume that the teleoperation task is a complex of $s$ serial subtasks $\mathcal{T} = \{\mathcal{T}_1,\,...,\,\mathcal{T}_s \}$. The \textit{task completion time} is given by $T_{exec} = \sum_{i=1}^{s}{T_i}$, 
where $T_i$ is the execution time of the $i$-th sub-task $\mathcal{T}_i$. This indicates how temporal demanding the whole teleoperation task is.
This total time should also include the system setup time and the associated UI's learning curve complexity. Indeed, the time an expert spends to setup a teleoperation system (denoted by $T_{setup}$) is of special importance in scenarios that need fast \textit{preparation time} (deployment time). In addition, an interface's \textit{learning curve} is evaluated in a supervised manner before the real task is started. 
More specifically, during learning, the UI's expert explains the usability, functionalities, robotic module's behaviours, and motion generation mechanism to the user. Afterwards, the user starts a learning trial to control the follower robot with the introduced UI in practice. When the user shows an accepted level of proficiency, the supervisor finishes the pre-training session and records the learning time that is denoted by $T_{learn}$.

The {subjective assessment tools}, on the other hand, evaluate the effectiveness and performance of a system, and in turn rate a user's perceived workload. One of the most used tools in this regard is the NASA-TLX that assesses workload based on mental demand (MD), physical demand (PD), temporal demand (TD), performance (PE), effort (EF), and frustration (FR). Each aspect is scored from 1 to 21, indicating the scale's strength with low, medium, and high labels. 

%% file: Sections/exp_setup.tex
\section{Experimental setup}
\label{section:experiments}
\begin{figure*}[!t]
     \centering
     \includegraphics[trim=0.15cm 0.7cm 0.15cm 0.15cm, width=0.975\linewidth]{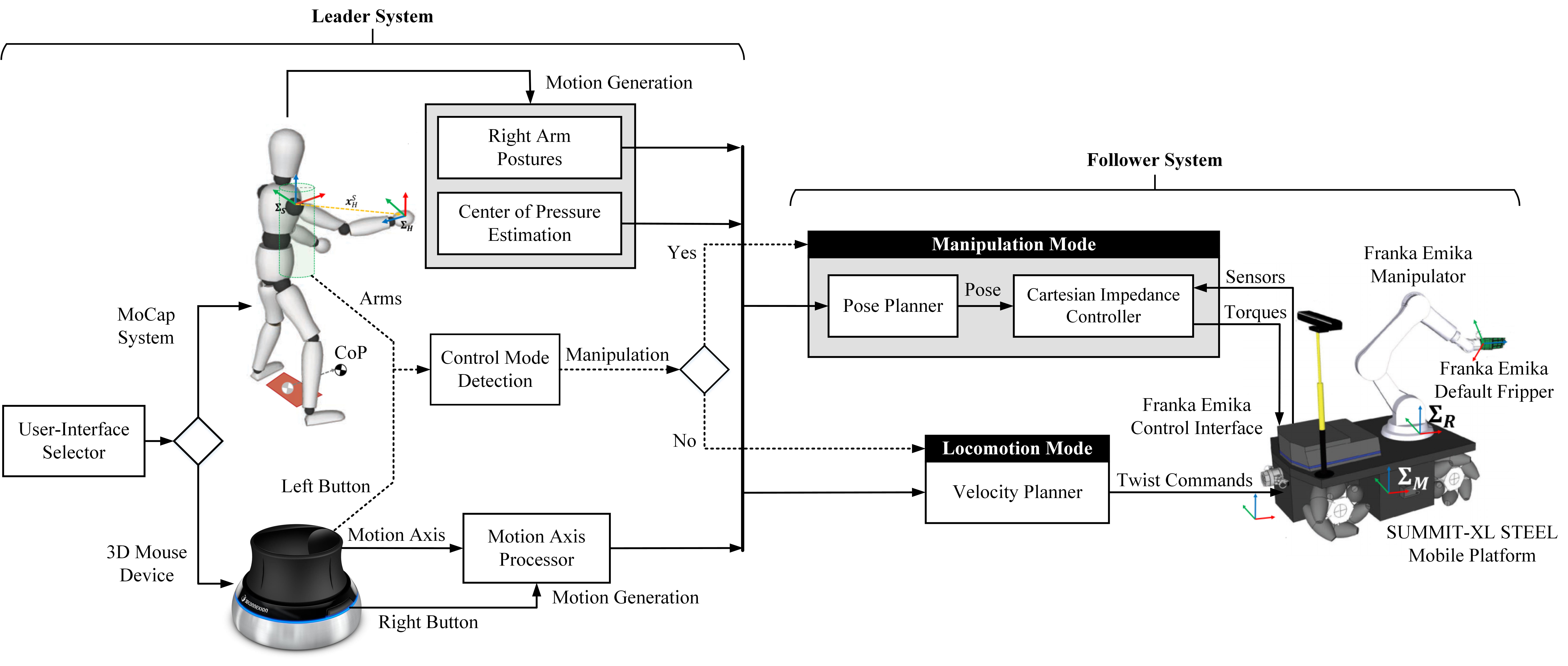}
     \caption{
     The  unified block  diagram  of  the  two teleoperation interfaces, i.e., MoCap system \cite{MocaTeleop} and 3D mouse device \cite{tro-vir}, for controlling a mobile robotic manipulator \cite{ott2008cartesian, albu2003cartesian, khatib1987unified,MocaTeleop}.}
    \label{fig:control_structure}
    \vspace{-0.1cm}
\end{figure*}
In this section, the experimental setup for the evaluation of the suggested framework is introduced. The setup includes two UIs (leader systems) to control a follower robotic system. The UIs are implemented based on
(i) a whole-body MoCap system in standing body posture, and 
(ii) a 3D mouse device in seated body posture. 
The choice of these two UIs is due to (i) the diversity among them, which allows a more comprehensive analysis of the proposed framework and (ii) the familiarity we have with their implementation and usage since they have been investigated and employed in previous studies \cite{MocaTeleop,tro-vir}. However, other different UIs may have been potentially employed to assess the developed metrics. Moreover, the follower robot is a mobile collaborative manipulator, controlled in two different modes, i.e., manipulation and locomotion. 

\begin{figure*}[!t]
     \centering
     \includegraphics[trim=0.0cm 0.0cm 0.0cm 0.0cm, width=1\linewidth]{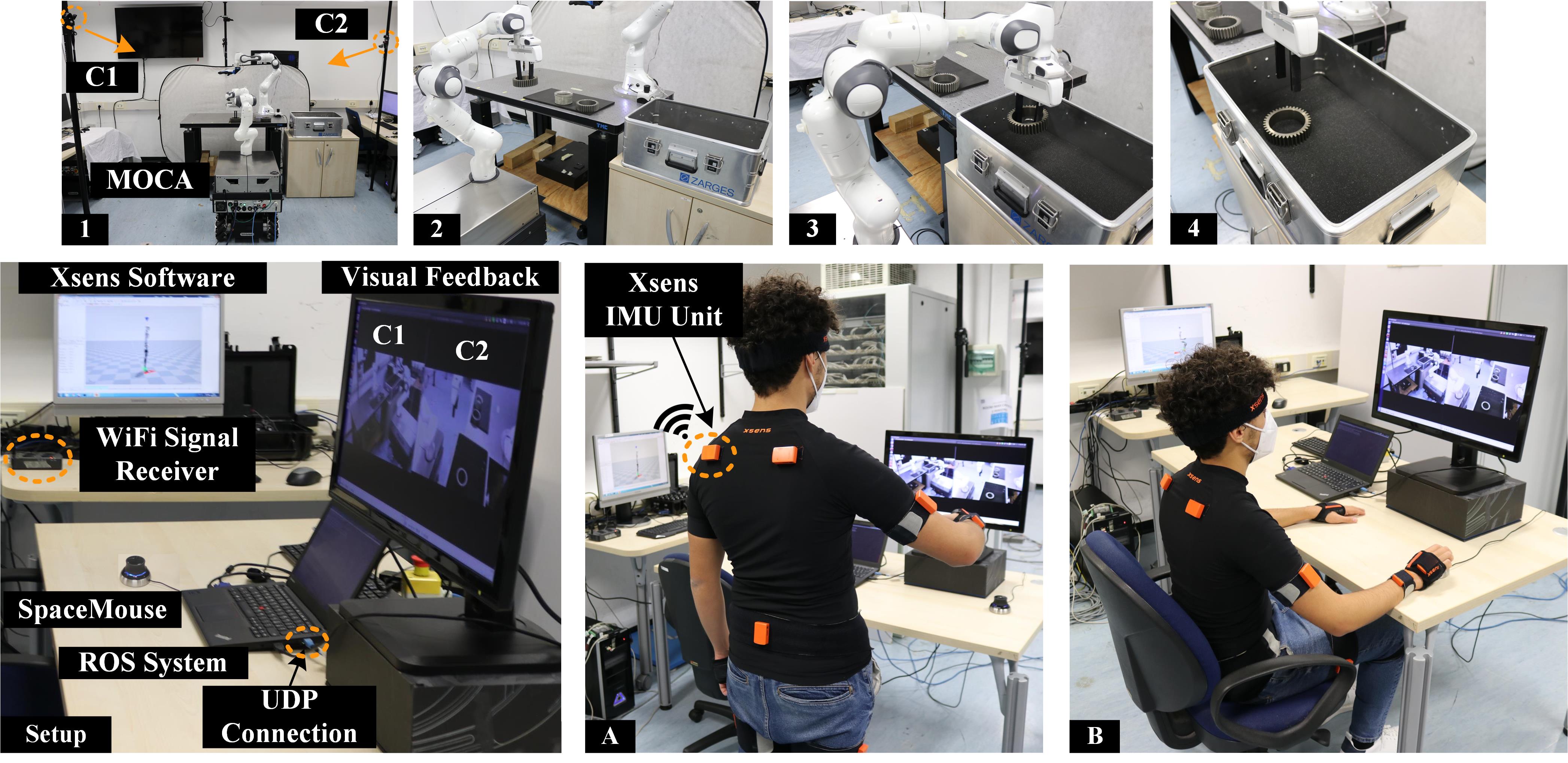}
     \caption{The snapshots of the task and experimental setup. First row: 
     (1) The user teleoperated MOCA to approach the first object by means of visual feedback. The installed cameras are labeled with C1 and C2, respectively.
     (2) The user grasped the object and navigated the robot toward the other table.
     (3) The user found the appropriate location to place the object inside the box (by using the second viewpoint). 
     (4) the object was placed in the box. Second row: 
     (Setup) the experimental elements used during the trials.
     (A) the standing posture with the MoCap system, 
     (B) the seated posture with the 3D mouse.
     }
    \label{fig:exp_setup}
    \vspace{-0.2cm}
\end{figure*}

In what follows, first, the follower robot system and its embedded loco-manipulation control structure are introduced. Then, the UIs and the strategy for generating the follower robot's reference trajectories are explained in detail. The overall system is illustrated in the block diagram of Fig.~\ref{fig:control_structure}. Whereas, the experimental setup is shown in  Fig.~\ref{fig:exp_setup}

\subsection{Follower mobile manipulator}
\label{follower_robot}
The MObile Collaborative Robotic Assistant (MOCA), which is an integration of a Robotnik{\textregistered} SUMMIT-XL STEEL mobile platform and a 7 DoF Franka Emika robot manipulator, is utilised as the follower mobile manipulator in our experiments. Moreover, we use the default Franka Emika's gripper to grasp the desired objects of the task. 
As illustrated in Fig.~\ref{fig:control_structure}, two separate \textit{control modes}, i.e., locomotion and manipulation, are considered to control the robot's whole-body loco-manipulation behaviour. These are toggled by the ``control mode detection" block triggered by a set of pre-defined patterns defined in the employed teleoperation UI. 

\subsubsection{Control mode: manipulation}
For the manipulation control mode, a ``Cartesian impedance controller" suitable for the robotic manipulators with rigid-joints and kinematically redundant structures is implemented \cite{ott2008cartesian, albu2003cartesian, khatib1987unified}. The virtual equilibrium point of this controller is generated by the ``pose planner" block in Fig. \ref{fig:control_structure} based on the output motion commands of the employed UI.
In the experiments, based on a trade-off between the interaction requirements with the stiff objects of the task and the required tracking performance during the teleoperation, the numerical values of the controller's parameters are selected as follows: the translational and rotational stiffness (critical damping) values are set to $300.0 \,\si{\newton\per\meter}$ ($2\,\sqrt{300.0}$) and $30.0\,\si{\newton\meter\per\radian}$ ($2\,\sqrt{30.0}$), respectively. Regarding the nullspace behaviour, the stiffness and damping values are chosen as
${K}_{\mathrm{n}} = 10.0\,\bm{I}_7$ and 
${B}_{\mathrm{n}} = 2\,\sqrt{10.0}\,\bm{I}_7$, respectively. 
Finally, the controller update rate is $1.0 \, \si{\kilo}\si{\hertz}$.

\subsubsection{Control mode: locomotion}
\label{subsec:locomotion}
In the locomotion control mode, on the other hand, a built-in low-level velocity controller is used to control the 2D pose 
of the omni-directional mobile base, equipped with four Mecanum wheels. The user's inputs are mapped to proper twist commands $\bm{v}={[\bm{v}_x,\, \bm{v}_y,\, \bm{\omega}_z]}^T$ where $\bm{v}_x$,\, $\bm{v}_y$, and $\bm{\omega}_z$ are the robot's linear velocity along x and y axes, and its angular velocity around z axis, respectively. This is done in the ``velocity planner" block (Fig. \ref{fig:control_structure}). 
During this mode, the robot manipulator is being controlled autonomously by the Cartesian impedance controller without the human user intervention. 
The update rate of the controller is set to $300.0 \, \si{\hertz}$. The maximum values of the twist elements are ${\bm{v}_x}_{max} = {\bm{v}_y}_{max} = 0.20 \, \si{\meter\per\second}$, and ${\bm{\omega}_z}_{max} = 0.25 \, \si{\radian\per\second}$. 

\subsection{User-interface 1: whole-body MoCap system}
\label{UI_1}
A wearable inertial-based system, the Xsens MVN BIOMECH (Xsens Technologies BV, Enschede, Netherlands), is selected as the whole-body MoCap device in this work. Compared to the camera-based systems that suffer from occlusions, limited range of camera visibility as well as unreliable continuous pose estimation and low update rate (e.g., when employing the OpenPose framework \cite{martinez2019openpose}), the inertial-based ones ensure the least amount of data loss (poor estimations) during the teleoperation tasks.
The MoCap system is used to satisfy two independent requirements: (i) tracking the body kinematic information to calculate the proposed metrics (Section \ref{section:metrics}) and (ii) generating the desired reference trajectories for the follower robot within the employment of the first UI (MoCap system). 
Thus, the Xsens Inertial Measurement Units (IMUs) and the corresponding WiFi-signal receiver module construct the ``hardware" block in Fig. \ref{fig:block_diagram}. The received signal is processed by the Xsens software bundle (Microsoft{\textregistered} Windows{\textregistered} $7.0$) and are sent to the ROS master (ROS Kinetic Kame in Ubuntu\textregistered  $16.04$) through the UDP networking protocol with the frequency of $60.0\, \si{\hertz}$. These are encapsulated in the ``software" block. 
The output signals are, then, used in the ``evaluation" block to meet the first requirement (metrics evaluation). On the other hand, and for the second requirement (motion generation), data is converted to the desired follower robot's reference trajectories by means of proper mapping functions. 
These functions are implemented in the ``right arm postures" and ``center of pressure estimation" blocks.

 In the locomotion mode, the inputs for the ``velocity planner" block are generated via the body Center of Pressure (CoP) movements projected on the 2D plane.
These are estimated by the ``Statically
Equivalent Serial Chain (SESC)" method \cite{kim2017real}. 
To avoid unwanted motions, a virtual polygon is assumed as a deadzone around the initial position of the CoP point on the floor (see the human manikin in Fig. \ref{fig:control_structure}). Consequently, body inclinations along x and y axes, out of the defined deadzone, $\delta{\bm{p}} \in \mathbb{R}^2$ generate the following virtual torques 
$\bm{\tau}_{v} = \bm{K}_{v}\,\delta{\bm{p}} +  \bm{B}_{v}\,\delta{\dot{\bm{p}}}$,
where, $\bm{K}_{v}$ and $\bm{B}_{v}$ are the virtual stiffness and damping matrices of the CoP movements model, respectively. Afterwards, $\bm{\tau}_{v}$ is processed by the designed admittance interface for the mobile base, and the outputs are sent to ``velocity planner" of the robot.
For the manipulation mode, on the other hand, the ``pose planner" receives the data from the ``right arm postures" block and, in turn, generates the required pose displacement for the "Cartesian impedance controller". 
This is done by utilising the 7 DoFs of the human right upper limb provided by the shoulder, elbow, and wrist joints. 
The switching between the control modes, here, is realised by the pre-defined arm gestures. When the user's arms are at his/her sides (N-pose), the locomotion mode is activated. Then, the user has to raise the right arm to trigger the manipulation mode. To switch back to the locomotion mode he/she first needs to raise the left arm and then to go back to the N posture. For detailed information, one may refer to \cite{MocaTeleop}.

\subsection{User-interface 2: 3D mouse device}
\label{UI_2}
The second interface implies the user to sit behind a desk \cite{tro-vir}. The operator controls the remote follower robot with a 3D mouse device, named SpaceMouse{\textregistered} Compact (3Dconnexion, UK). This mouse has a 6 DoF motion sensor (motion axis) and two push buttons. For the purposes of our teleoperation tasks, these buttons are used to enable the operators to switch between modes. Indeed, the user can toggle between the control modes by pressing the left push button and change the motion modes (i.e., translation and rotation) with the right button. The switch between motion modes, however, is just available in the manipulation mode. This is because the dimension of the task-space is considered to be 6, indicating that all the mouse's DoFs are required for controlling the manipulator's end-effector.

Regarding the motion generation for the follower robot, initial tests proved that the mouse's DoFs are highly coupled and sensitive. This makes the generation of precise and decoupled reference trajectories near impossible. Thus, the reference trajectories are generated after processing the raw mouse's motion axis information (DoFs) in the ``motion axis processor" block (Fig. \ref{fig:control_structure}).
More specifically, the mouse's raw data are, first, read and stored as $\delta \bm{x}_{raw} ={[\delta \bm{p}^T_{raw},\,\delta \bm{\epsilon}^T_{raw}]}^T$ with the frequency of $50.0\, \si{\hertz}$. $\delta \bm{p}_{raw} \in {\mathbb{R}}^{3}$ and $\delta \bm{\epsilon}_{raw} \in {\mathbb{R}}^{3}$ are the translational and rotational displacements along and around the mouse's motion axis, respectively. Then, these are normalized to a signed percentage values, denoted by $\delta \bm{x}^\dagger_{raw}$. Afterwards, the motion mode is checked (just in the manipulation control mode), assigning $\delta \bm{\epsilon}^\dagger_{raw}$ or $\delta \bm{p}^\dagger_{raw}$ to $\bm{0}$ when the translation or rotation motion mode are activated, respectively.
Next, a moving average filter is applied with a window of size $N$, populated with the current $\delta \bm{x}^\dagger_{raw}$ and its last $N-1$ values. The average value of each motion axis $j$ of the mouse over the past $N$ samples is defined as ${\delta \bar{\bm{x}}_j}(k)$ at time instant $k$.
Finally, the maximum value of ${\delta \bar{\bm{x}}_j}$ and its corresponding axis label $j^{\star}$ are fetched and the other values are set to  $\bm{0}$. Consequently, the desired displacement vector is generated as $\delta {\bm{x}}_d={[\delta \bm{p}^T_{d},\,\delta \bm{\epsilon}^T_{d}]}^T$, where only the $j^{\star}$-th element is non-zero. In the locomotion mode, $\delta {\bm{p}}^x_d$, $\delta {\bm{p}}^y_d$, and $\delta {\bm{\epsilon}}^z_d$ are converted to the desired twist command of the platform based on ${\bm{v}_x}_{max}$, ${\bm{v}_y}_{max}$, and ${\bm{\omega}_z}_{max}$ (see Section \ref{subsec:locomotion}). For the manipulation mode, on the other hand, all of the elements may be used based on the motion mode and the pre-set maximum Cartesian steps of each DoF (for translational and rotational motions, these steps are set to $0.01\, \si{\meter}$ and  $0.5\, \si{\deg}$, respectively).

%% file: Sections/exp_validation.tex
\section{Experimental Evaluation}
\label{section:exp_validation}
\subsection{Task description}
To evaluate the proposed framework, we asked eleven healthy subjects of different ages ($27.54\, \pm\, 3.09$ years), body heights ($178.95\, \pm\, 5.39$ cm) and genders (nine males and two females) to execute an identical task by using the introduced UIs (sections \ref{UI_1} and \ref{UI_2}). Thus, each subject performed the task in two independent trials (groups). More specifically, the trials carried out by means of the MoCap system and 3D mouse device are labeled as ``trial 1" and ``trial 2", respectively. The first trial was followed by the second one after a ten-minute break. The experimental setups related to these trials are shown in the second row of Fig.~\ref{fig:exp_setup}. 
The whole experimental procedure was conducted in accordance with the Declaration of Helsinki and the protocol was approved by the regional ethics committee of Liguria (Protocol IIT-HRII-ERGOLEAN, 156/2020, DB-id 10215).

Regarding the task, a remote pick-and-place operation with industrial gear pieces was considered to be executed by the participants. The task's snapshots are displayed in the first row of Fig.~\ref{fig:exp_setup}. The latter illustrate the sequence of sub-tasks that should be followed one after another, for picking one of the task's target objects and placing it into a desired box. 
In each trial (i.e., the MoCap or the 3D mouse), the subject started moving the follower mobile manipulator from the initial location toward a table on which three different industrial gears were placed (Fig. {\ref{fig:exp_setup}}-1). There, she/he grasped the first object (Fig. {\ref{fig:exp_setup}}-2) and moved the robot near to an empty box located on another table (Fig. {\ref{fig:exp_setup}}-3). Finally, she/he placed the object into the box (Fig. {\ref{fig:exp_setup}}-4). This cycle was repeated for the middle and right objects, respectively. 
For the visual feedback, a monitor display was used to show the remote environment to the user with two different viewpoints. Indeed, two RGB cameras were installed in the remote environment whose locations were optimised based on the task's requirements. The location of the installed cameras are shown by ``C1" and ``C2" labels in Fig. \ref{fig:exp_setup}-1, and corresponding viewpoints of the remote environment are displayed in Fig. \ref{fig:exp_setup}-setup. 

\subsection{Analyses}
In this section, numerical, graphical, and statistical data analyses over the performed experiments are discussed.

\begin{figure}[!t]
     \centering
     \includegraphics[trim=0.1cm 0.1cm 0.1cm 0.1cm, width=0.825\linewidth]{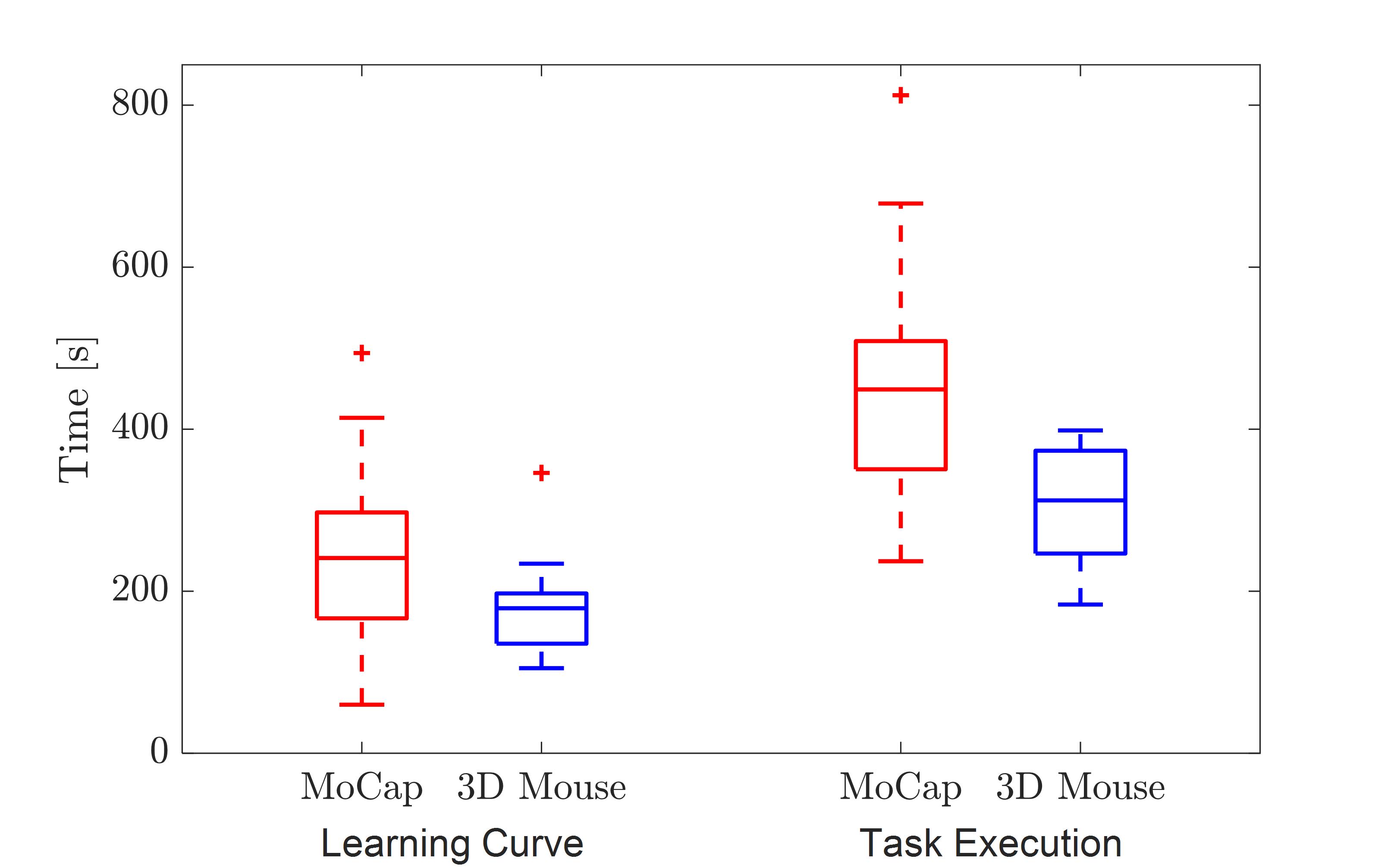}
     \caption{
     Descriptive statistics by means of the boxplot method to compare the employed UIs in the sense of learning curve duration and task execution time.   
     ``+'' denotes the outlier data. 
     }
     \label{fig:time_box_plots}
\end{figure}

Starting with the time-related metrics, the median ($\mathrm{M}$) and InterQuartile Range ($\mathrm{IQR}$) of the data, among the eleven subjects, are chosen as the statistical parameters. 
For the learning curve, we have: $\mathrm{M_{MoCap}} = 241.00\, \si{\second}$ 
and 
$\mathrm{IQR_{MoCap}} = 130.75\, \si{\second}$, 
and 
$\mathrm{M_{3D}} = 179.00\, \si{\second}$ 
and 
$\mathrm{IQR_{3D}} = 62.00\, \si{\second}$. For the execution time: 
$\mathrm{M_{MoCap}} = 449.08\, \si{\second}$ and 
$\mathrm{IQR_{MoCap}} = 158.18\, \si{\second}$, and 
$\mathrm{M_{3D}} = 312.13\, \si{\second}$ and 
$\mathrm{IQR_{3D}} = 126.96\, \si{\second}$.
The results suggest the UI with the 3D mouse device was easier to be learned by the participants ($25.72\%$ reduction in the learning time). Also, the task was executed $30.49\%$ faster with this UI.
These results are shown in the boxplots of Fig. \ref{fig:time_box_plots}.

\begin{figure}[!t]
     \centering
     \includegraphics[trim=0.1cm 0.3cm 0.1cm 0.3cm, width=0.825\linewidth]{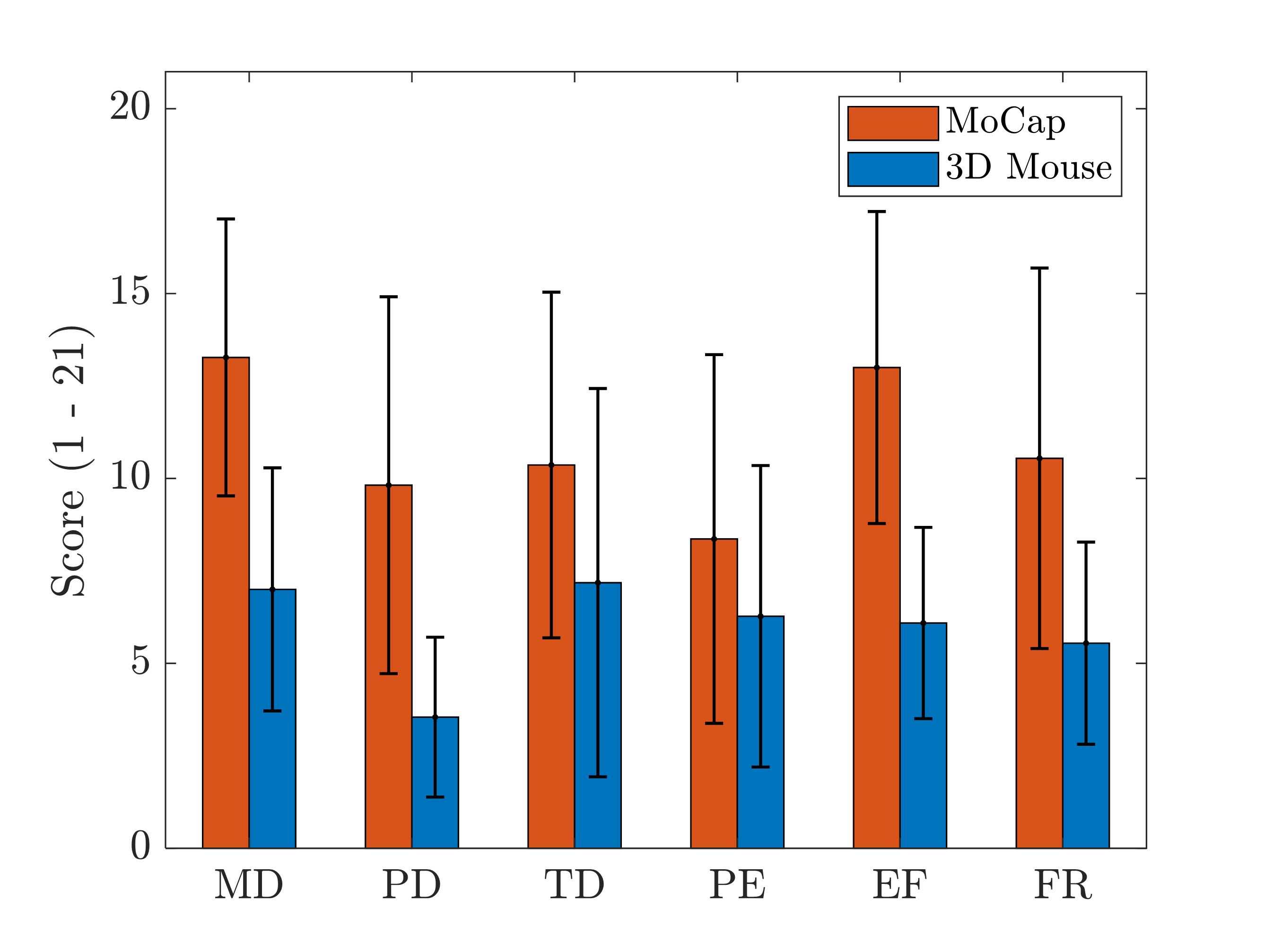}
     \caption{NASA-TLX results for the experimental groups, i.e., the MoCap and 3D mouse UIs. 
     }
    \label{nasa_tlx}
\end{figure}

\begin{table}[!t]
\caption{The mean and standard deviation values of the NASA-TLX assessment tool among eleven subjects.}
\label{table:nasa}
\begin{adjustbox}{width=0.475\textwidth}
\begin{tabular}{@{}lcccccc@{}}
\toprule
\textbf{UI}       & \textbf{MD}      & \textbf{PD}                                                & \textbf{TD}      & \textbf{PE}     & \textbf{EF}     & \textbf{FR}    \\ \midrule
\textbf{MoCap}    & $13.27 \pm 3.74$ & \begin{tabular}[c]{@{}l@{}}$9.81 \pm 5.09$\end{tabular} & $10.36 \pm 4.67$ & $8.36 \pm 4.98$ & $13.00 \pm4.21$ & $10.54\pm5.14$ \\
\textbf{3D Mouse} & $7.00\pm 3.28$   & $3.54\pm 2.16$                                             & $7.18\pm 5.25$   & $6.27\pm 4.07$  & $6.09\pm 2.58$  & $5.54\pm 2.73$ \\ \bottomrule
\end{tabular}
\end{adjustbox}
\end{table}

The results of the subjective NASA-TLX questionnaire are  displayed  in  Fig. \ref{nasa_tlx}.  The exact mean and standard deviation values related to this questionnaire, among the eleven subjects, are reported in Table. \ref{table:nasa}. 
The  participants  showed  less  mental, physical, and temporal demands with the 3D mouse UI (MD, PD, and TD, respectively). Especially, this difference is more evident in the physical and mental scales. Instead,  the ``performance" scale is quite similar in both groups with a slight lower score for the MoCap UI (here, $1.0$ and $21.0$ indicate ``perfect" and ``failure", respectively). 
Moreover, and as it was expected, the perceived ``effort" is way higher in the first group, most likely due to the standing nature of the employed UI. Finally, the ``frustration" scale is lower in the second group, which is probably because of the difference between the mapping functions of the employed motion generation interfaces.

\begin{table*}[!t]
\centering
\caption{The normalized numerical comparison between the studied groups (teleoperation UIs) in the experiments. In the table cells, ``$\cdot,\,\cdot$" denote the median and $\mathrm{IQR}$ pair, i.e., ``$\mathrm{M},\, \mathrm{IQR}$". }
    \begin{adjustbox}{width=0.9\textwidth}
\begin{tabular}{@{}c|c|ccc|ccc@{}}
\toprule
\multicolumn{2}{c}{}                                                                          & \multicolumn{3}{c}{\textbf{Group 1: MoCap}}                           & \multicolumn{3}{c}{\textbf{Group 2: 3D Mouse}}
\\ \midrule
\textbf{Part}                                                                & \textbf{Joint/Link} & \textbf{Posture comfort} & \textbf{Joints' usage} & \textbf{RoM comfort}        & \textbf{Posture comfort} & \textbf{Joints' usage} & \textbf{RoM comfort}        \\ \midrule
\multirow{3}{*}{\begin{tabular}[c]{@{}c@{}}vertebral \\ column\end{tabular}} & head           & $0.0561,\, 0.0150$     & $0.0501,\, 0.0607$  & $0.8794,\, 0.0702$ & $0.0528,\, 0.0240$     & $0.0619,\, 0.0407$  & $0.8554,\, 0.1029$ \\
                                                                             & neck           & $0.0494,\, 0.0080$     & $0.0266,\, 0.0328$  & $0.9959,\, 0.0084$ & $0.0311,\, 0.0153$     & $0.0331,\, 0.0216$  & $0.9882,\, 0.0120$ \\
                                                                             & pelvis         & $0.0267,\, 0.0053$     & $0.0465,\, 0.0325$  & $0.7618,\, 0.0111$ & $0.0199,\, 0.0104$     & $0.0075,\, 0.0187$  & $0.7870,\, 0.0170$ \\ \midrule
\multirow{3}{*}{\begin{tabular}[c]{@{}c@{}}left upper \\ limb\end{tabular}}  & shoulder       & $0.0520,\, 0.0552$     & $0.0879,\, 0.0764$  & $0.7975,\, 0.0187$ & $0.0348,\, 0.0378$     & $0.0252,\, 0.0155$  & $0.8196,\, 0.0102$ \\
                                                                             & elbow          & $0.1032,\, 0.0723$     & $0.1613,\, 0.1550$  & $0.7060,\, 0.0346$ & $0.0226,\, 0.0149$     & $0.0353,\, 0.0142$  & $0.7644,\, 0.0201$ \\
                                                                             & wrist          & $0.1365,\, 0.0497$     & $0.0684,\, 0.0848$  & $0.8334,\, 0.0168$ & $0.0254,\, 0.0164$     & $0.0167,\, 0.0315$  & $0.8776,\, 0.0268$ \\ \midrule
\multirow{3}{*}{\begin{tabular}[c]{@{}c@{}}right upper \\ limb\end{tabular}} & shoulder       & $0.1147,\, 0.0691$     & $0.2614,\, 0.1126$  & $0.6953,\, 0.0974$ & $0.0383,\, 0.0248$     & $0.0287,\, 0.0369$  & $0.8066,\, 0.0241$ \\
                                                                             & elbow          & $0.5033,\, 0.1178$     & $0.5808,\, 0.3809$  & $0.4552,\, 0.0527$ & $0.0581,\, 0.0359$     & $0.0454,\, 0.0605$  & $0.7449,\, 0.0460$ \\
                                                                             & wrist          & $0.6683,\, 0.1923$     & $0.1999,\, 0.1548$  & $0.8144,\, 0.0521$ & $0.0663,\, 0.0544$     & $0.0658,\, 0.0400$  & $0.8707,\, 0.0121$ \\ \midrule
\multirow{3}{*}{\begin{tabular}[c]{@{}c@{}}left lower \\ limb\end{tabular}}  & hip            & $0.0268,\, 0.0045$     & $0.0387,\, 0.0389$  & $0.7258,\, 0.0034$ & $0.0189,\, 0.0093$     & $0.0089,\, 0.0148$  & $0.7233,\, 0.0102$ \\
                                                                             & knee           & $0.0186,\, 0.0124$     & $0.0342,\, 0.0219$  & $0.6529,\, 0.0100$ & $0.0173,\, 0.0141$     & $0.0052,\, 0.0038$  & $0.6695,\, 0.0039$ \\
                                                                             & ankle          & $0.0137,\, 0.0039$     & $0.0320,\, 0.0263$  & $0.6793,\, 0.0030$ & $0.0178,\, 0.0178$     & $0.0053,\, 0.0024$  & $0.6743,\, 0.0066$ \\ \midrule
\multirow{3}{*}{\begin{tabular}[c]{@{}c@{}}right lower \\ limb\end{tabular}} & hip            & $0.0266,\, 0.0068$     & $0.0518,\, 0.0405$  & $0.6959,\, 0.0114$ & $0.0200,\, 0.0062$     & $0.0086,\, 0.0201$  & $0.7096,\, 0.0139$ \\
                                                                             & knee           & $0.0192,\, 0.0033$     & $0.0313,\, 0.0226$  & $0.6553,\, 0.0088$ & $0.0202,\, 0.0134$     & $0.0058,\, 0.0060$  & $0.6714,\, 0.0040$ \\
                                                                             & ankle          & $0.0157,\, 0.0055$     & $0.0371,\, 0.0351$  & $0.6802,\, 0.0102$ & $0.0237,\, 0.0155$     & $0.0046,\, 0.0050$  & $0.6733,\, 0.0019$ \\ \bottomrule
\end{tabular}
\label{table: numerical_comp}
\end{adjustbox}
\end{table*}

\begin{figure*}[!t]
\centering
\subfloat[Metric: posture comfort.]{%
  \includegraphics[trim=0cm 0cm 0cm 0cm,clip,width=0.9\textwidth]
  {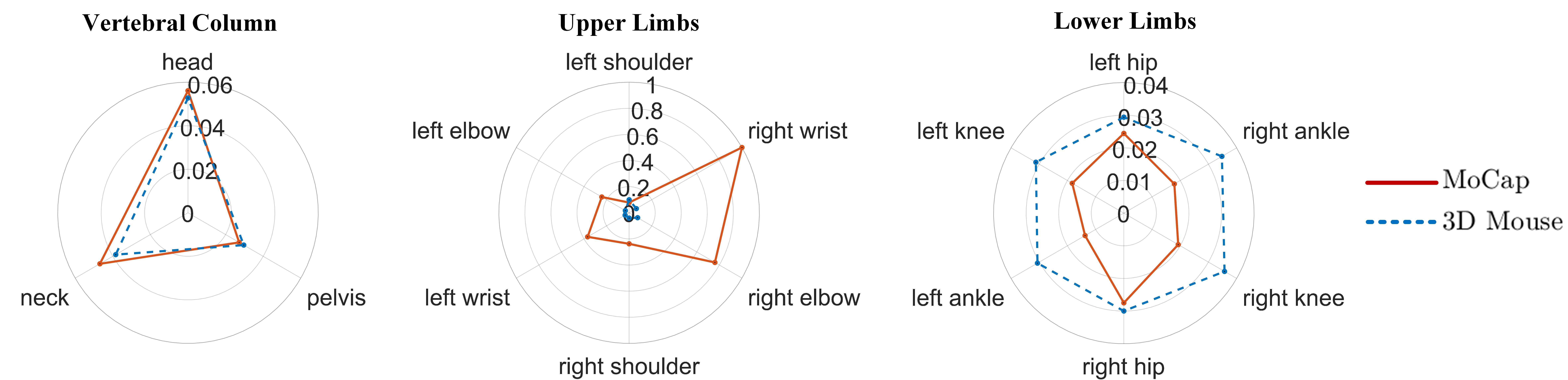}%
  \label{subfig:idx1}
}\\
\subfloat[Metric: Range of Motion (RoM) comfort.]{%
  \includegraphics[trim=0cm 0cm 0cm 0cm,clip,width=0.9\textwidth]
  {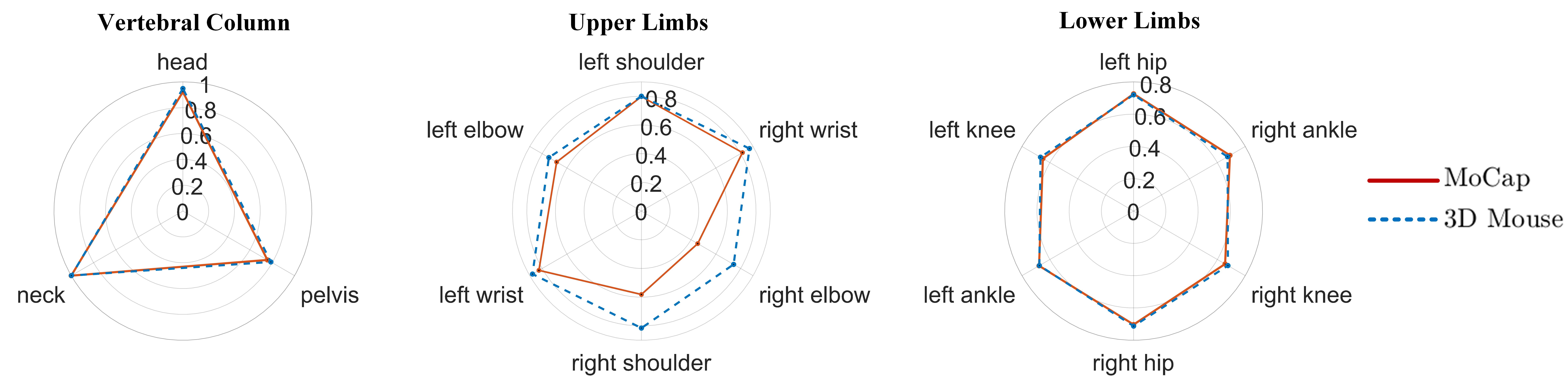}%
  \label{subfig:idx2}
}\\
\subfloat[Metric: joints' usage.]{%
  \includegraphics[trim=0cm 0cm 0cm 0cm,clip,width=0.9\textwidth]
  {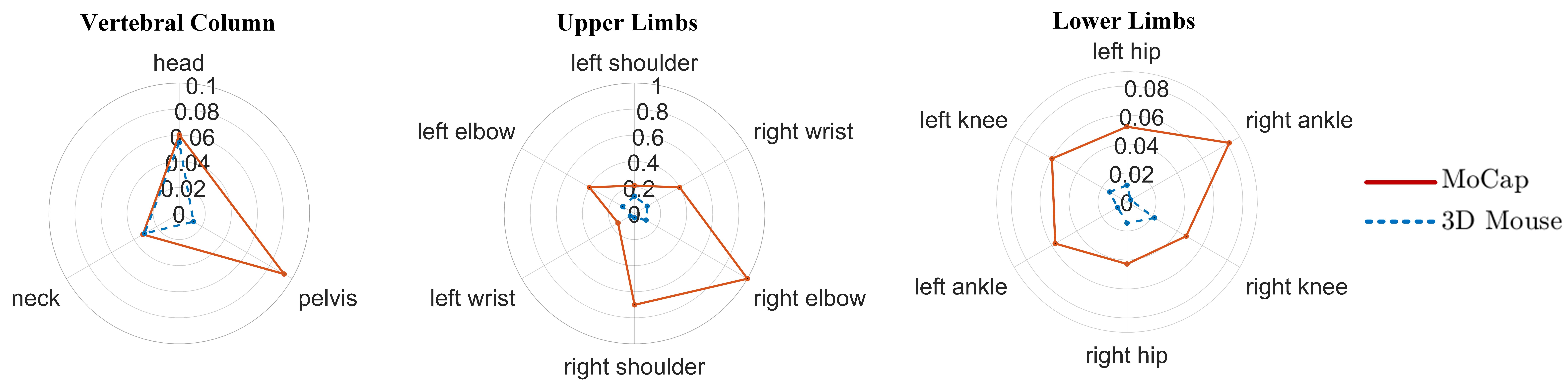}%
  \label{subfig:idx3}
}
\caption{The normalized polar-plot representations of the introduced quantitative metrics. The results are related to the 8-th subject.}
\label{fig:polar_plot}
\end{figure*}

From the quantitative perspective, normalized whole-body comparisons are carried out on the studied groups based on the proposed metrics in Section \ref{section:metrics}. As a result,
the statistical parameters, formatted as ``$\mathrm{M},\, \mathrm{IQR}$" pair, computed among eleven subjects, are listed in Table \ref{table: numerical_comp} for each body part.
Besides, a graphical representation of the indexes is provided in Fig. \ref{fig:polar_plot} by means of the polar plots. In this figure, each row illustrates a particular index, being the whole-body divided to three sections: ``vertebral column", ``upper limbs", and ``lower limbs". To avoid clutter, we just consider one of the subjects (8-th subject). By calculating the corresponding polynomial areas, a ratio is obtained for each body part giving the relative strength of the 3D mouse UI over the MoCap one, or vice versa. These results are discussed in what follows.

For the ``posture comfort" index, normalized average values of \eqref{optimal_pose} during the task's execution time are calculated for each body part. It should be noted that the users started the task in a \textit{pre-defined ideal body posture} approved by the task supervisor. So, the divergence from this initial posture is served as the first metric. 
Results show remarkable divergence values in the body upper limbs when the user is equipped with the MoCap UI. 
However, the corresponding values for the vertebral column and the lower limbs are not considerably different. These are also shown graphically for one of the subjects in Fig. \ref{subfig:idx1}. Compared with the other body parts, the greatest divergence from the ideal posture is noticed in the right upper limb in the first trial (the MoCap UI). The lower limbs, on the other hand, have a better condition in this trial for this particular user. Indeed, the ratio values (by calculating the polynomial areas) for the vertebral column, upper limbs, and lower limbs are $1.15$, $60.17$, and $ 0.39$, respectively.

As we discussed earlier, the more the values of \eqref{metric_rom} are close to 1, the more comfort the body experiences in terms of RoM. Based on the data presented in the table, the users performed the task with a totally comfortable posture while using the second UI. Indeed, the minimum median value for this setup is $\mathrm{M}=0.6695$ which is an acceptable value for the RoM index (the threshold is set to $70.0\%$). The same conclusion holds also for the first UI but with an exception. Here, the right elbow is exposed to a non-comfortable posture during the task ($\mathrm{M}=0.4552$). It should be noted that the average values of \eqref{metric_rom} are normalized for the body parts throughout the task completion, and listed in the table. This is also shown for the studied subject in Fig. \ref{subfig:idx2}, and the corresponding ratios are $0.9586$, $0.7334$, and $0.9893$ for the vertebral column, upper limbs, and lower limbs, respectively. This shows a slight better ``RoM comfort" index status for the second UI compared to the first one.

Regarding the normalized ``joints' usage" values in the table, these are calculated by the numerical integration of \eqref{metric_joints_usage} via the trapezoidal method during the task. As expected, body joints were used more by the subjects when they employed the first UI; especially the lower and upper limbs. However, the neck and head joints' usage have a negligible difference. Moreover, the graphical results in \ref{subfig:idx3} yield the following ratios for the vertebral column, upper limb, and lower limbs, respectively: $3.84$, $29.38$, and $24.14$. As it may be seen in this figure, the most employed body joints are the right elbow and shoulder while the user is performing the task with the first UI.

\begin{figure}[!t]
     \centering
     \includegraphics[trim=0.1cm 0.1cm 0.1cm 0.1cm, width=0.95\linewidth]{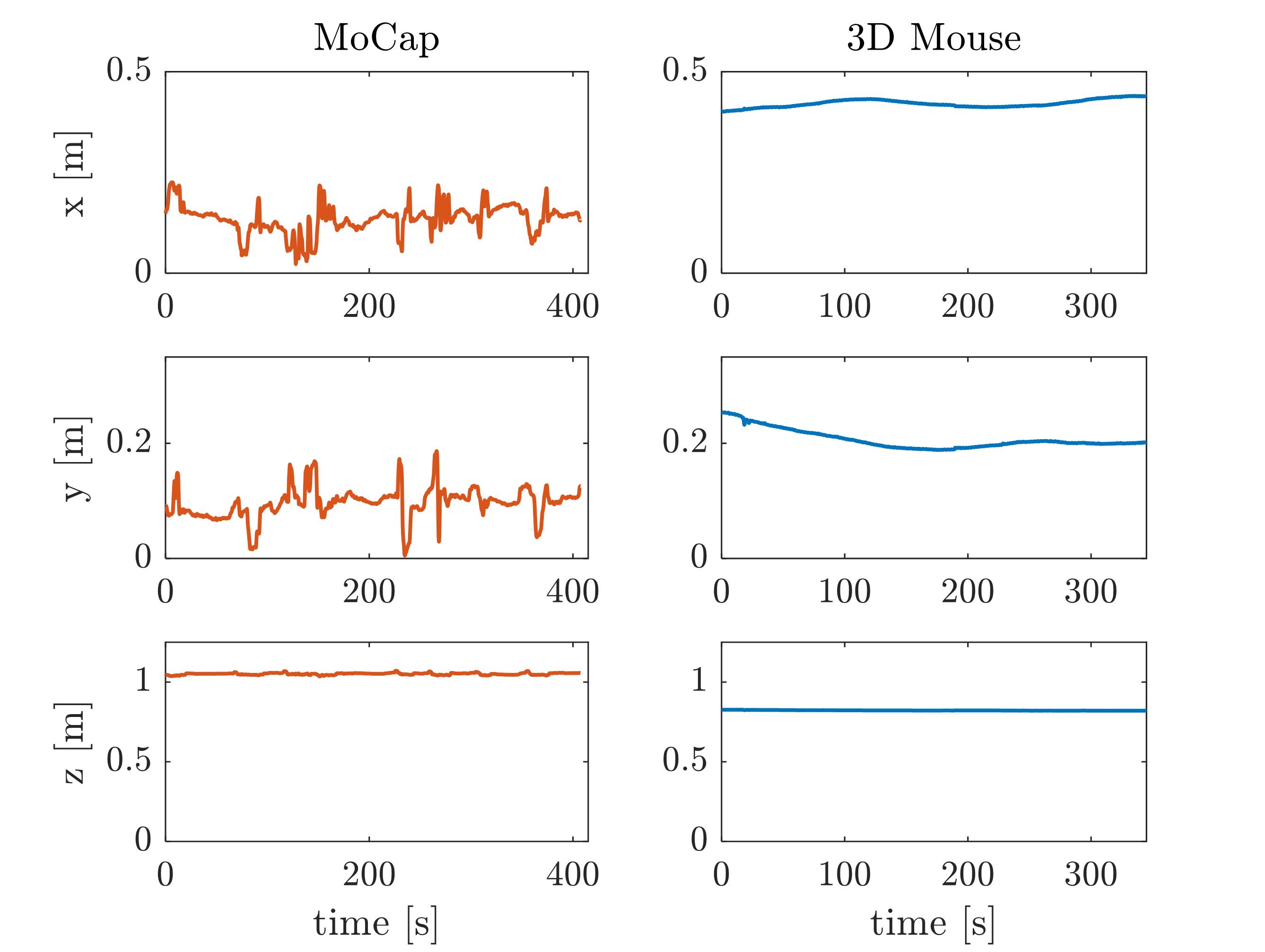}
     \caption{The evolution of body Center of Mass (CoM) parameter in time. The results are related to the 8-th subject. }
     \label{fig:CoG}
\end{figure}

Furthermore, the  statistical analyses of the body ``CoM divergence" is investigated in this section. The median and $\mathrm{IQR}$ values of the normalized index in \eqref{metric_cog} (divergence from the average point) over the eleven subjects are:
$\mathrm{M_{MoCap}} = 0.5978$ and 
$\mathrm{IQR_{MoCap}} = 0.4175$ for the MoCap UI, and 
$\mathrm{M_{3D}} = 0.1816$ and 
$\mathrm{IQR_{3D}} = 0.1301$ 
for the 3D mouse one. 
Moreover, the body CoM information related to the afore-mentioned subject during both trials are shown in Fig. \ref{fig:CoG}. As it may be seen, the body CoM experienced more fluctuations during the teleoperation task when the user employed the MoCap UI. This is mostly because of the whole-body motion generation technique used within this UI, which implies the user to generate the navigational motions with the lateral body CoM motions. These fluctuations, on the other hand, are not seen in the seated UI and the body CoM stayed within a limited region during the task and this contributes to less perceived effort in the human body. In addition, the numerical integration of the suggested index in \eqref{metric_cog} (divergence from the average point) via the trapezoidal method yields the ratio of $2.60$ between the MoCap and 3D mouse UIs. 

%% file: Sections/conclusions.tex
\section{Conclusion and future perspectives}
\label{section:conclusion}
In this work, we developed a quantitative physical ergonomics assessment framework based on a set of human kinematics-related information. The objective was to create a tool that can evaluate and compare 
different types of teleoperation UIs so that the system designers can quantify the impact of UI elements on the overall teleoperation system's performance and ergonomics. Experiments with eleven participants in using two different teleoperation UIs, i.e., standing posture interface with a whole-body MoCap system and the seated one with a 3D mouse device, showed the proposed framework's effectiveness. 

The effects of well-known teleoperation factors such as communication time-delays and vision feedback (i.e., the information about the remote environment provided by 2D cameras) were not investigated in this work. 
The former holds true because of the local network used throughout the experiments, while the latter imposes the same effects to both UIs. In addition, the two UIs implemented a unilateral teleoperation architecture, thus the stability issues of the bilateral force-reflecting architectures were not of our interest. 

Future works will focus on the enhancement of the ergonomics assessment. The effects of task repetition and frequency, which has been recognised as a significant contributor to musculoskeletal disorders, as well as users' experience on the proposed metrics will be investigated. 
Besides, our experiments are carried out by non-expert users while in real-world applications, teleoperation tasks are executed by operators who undergo long training sessions. Hence, the latter will be considered as subjects in future experimental sessions. 
Finally, alternative teleoperation UIs will be taken into account to validate the proposed evaluation framework.